\documentclass[amsmath,amssymb,aip,pof]{revtex4-1}

\usepackage{graphicx,amssymb,amsmath}
\usepackage{xcolor}
\usepackage[applemac]{inputenc}
\usepackage{epstopdf}
\usepackage{float} 

\begin{document}

\title{Investigation of resonances in gravity-capillary wave turbulence}
\author{Quentin Aubourg}
\affiliation{Universit\'e Grenoble Alpes, LEGI, CNRS  F-38000 Grenoble, France}
\author{Nicolas Mordant}
\email[]{nicolas.mordant@ujf-grenoble.fr}
\affiliation{Universit\'e Grenoble Alpes, LEGI, CNRS  F-38000 Grenoble, France}
\affiliation{Institut Universitaire de France, 103, bd Saint Michel, F-75005 Paris, France}

\begin{abstract}
We report experimental results on nonlinear wave coupling in surface wave turbulence on water at scales close to the crossover between surface gravity waves and capillary waves. We study 3-wave correlations either in the frequency domain or in wavevector domain. We observe that in a weakly nonlinear regime, the dominant nonlinear interactions correspond to waves that are collinear or close to collinear. Although the resonant coupling of pure gravity waves is supposed to involve 4 waves, at the capillary crossover we observe a nonlocal coupling between a gravity wave and 2 capillary waves. Furthermore nonlinear spectral spreading permits 3-gravity wave coupling. These observations raise the question of the relevance of these processes in the oceanographic context and in particular the range of frequencies of gravity waves that may be impacted. 
\end{abstract}

\maketitle

\section{Introduction}

The weak turbulence theory (WTT) has been developed since the 60's with initial motivations in oceanography and plasmas physics (see ~\cite{R1,R2,R3} for reviews). It provides a path from the dynamical wave equations to a statistical theory and notably the time evolution of the wave spectrum. This major step torward a statistical theory of a turbulent system can only be achieved under the hypotheses of both asymptotically large system and asymptotically weak nonlinearity. In this framework, a natural scale separation exists between the fast period of the oscillations of the waves and the slow nonlinear evolution due to coupling with the other waves. This separation permits to develop an asymptotic statistical theory. Because of the weakness of the nonlinear coupling only resonant wave can develop a long time cumulative coupling and in this way exchange significant amounts of energy. These waves verify the resonance conditions on wavenumbers $\mathbf k_i$ and angular frequencies $\omega_i$:
\begin{equation}
\mathbf k_1=\mathbf k_2+\mathbf k_3 \quad\quad \omega_1=\omega_2+\omega_3
\end{equation}
in the case of 3 interacting waves. The number of interacting waves to consider is determined generally by the order of the nonlinear term (3-wave coupling for quadratic nonlinearity, 4-wave coupling for cubic ones,...). Exceptions come from the curvature of the linear dispersion relation that sometimes prevents from finding any solutions for 3-wave interaction for instance despite quadratic non linear terms. This is the case for dispersion relations $\omega\propto k^\alpha$ for $\alpha<1$ as is the case for pure gravity surface waves for which $\omega=\sqrt{gk}$ ($g$ being the acceleration of gravity). In such a case 4-wave resonance have to be investigated. In real systems that have a finite size, discretization of frequencies and wavevectors renders the situation even more complex as the resonant sets can become extremely restricted ~\cite{Kartashova,Kartashova1,Kartashova2,Naz06}. In real systems, the magnitude of the waves and thus of the non linear coupling is also finite. Energy exchanges between waves decrease the coherence of the waves and thus induce a finite spectral width of the modes. The resonances are thus approximate:
\begin{equation}
\mathbf k_1\approx\mathbf k_2+\mathbf k_3 \quad\quad \omega_1\approx\omega_2+\omega_3
\end{equation}
as some uncertainty exists on the frequencies and the wavenumbers due to non linear spectral widening. Depending on the amplitude of the mismatch this set of approximate resonances is possibly much larger than that of exact resonances and this may enable a much more efficient wave coupling and thus much a more efficient energy transfer among waves.

The notion of wave resonance is thus at the core of the WTT. Not only it permits to derive the evolution equations for statistical quantities like spectra but solutions of these equations can be exhibited in stationary situations notably in the forced and thus out of equilibrium case ~\cite{R1,R2,R3}. The phenomenology associated to this latter case is that of the energy cascade in scale from the forced scales (usually large scales) to the scales at which dissipation dominates and ultimately removes energy into heat. The stationary out of equilibrium spectrum has been called Kolmogorov-Zakharov (KZ) spectrum in honor of V.E. Zakharov and because of the analogy with the Kolmogorov spectrum of Navier-Stokes turbulence. In the case of pure capillary waves (short wavelength limit of surface waves on a liquid) this spectrum of the surface deformation can actually be predicted by dimensional analysis (and confirmed analytically) to be $E_\zeta(k)\propto \gamma^{1/4}P^{1/2}/k^{7/4}$ where $k=||\mathbf k||$, $\gamma$ is the surface tension and $P$ is the energy injected (and dissipated) in the stationary system. The exponent $1/2$ comes from the 3-wave interaction. For pure gravity waves (long wavelength limit) the spectrum is $E_\zeta(k)\propto g^{1/2}P^{1/3}/k^{5/2}$, the coefficient $1/3$ being due to 4-wave resonances as discussed above. In the general case of capillary-gravity waves i.e. for wavelengths of centimeter scale, the dispersion relation is
\begin{equation}
\omega=\sqrt{gk+\frac{\gamma}{\rho} k^3}\, .
\label{rd}
\end{equation}
As it is not a pure power law the analytic derivation of the KZ spectrum is much more difficult. It has been shown ~\cite{Mcgoldrick1965,Simmons1969} that for any given value of $\omega_2$ 3-wave resonances are possible but that the interaction becomes quite non-local when it takes values in the gravity range. Preliminary examinations of resonances in the turbulent state have suggested that the interactions are much more efficient in the almost unidirectional situation in which all the wave vectors are close to be collinear ~\cite{Aubourg2015}.

In this article we investigate in more details the wave coupling in the turbulent state. We investigate 3-wave correlations either in the frequency or the wavenumber domain. This statistical estimator permits to check the existence of wave resonances. We study two regimes: a weakly non linear regime and a regime of stronger forcing. The experimental setup is described in part II. General spectral characteristics of the wave field are presented in part III. The resonances are specifically studied in part IV: first the theoretical exact resonances, then correlations in frequency domain or in wavevector space. Conclusions are drawn in the last part.

\section{Experimental setup}

\begin{figure}[!htb]
\includegraphics[width=14cm]{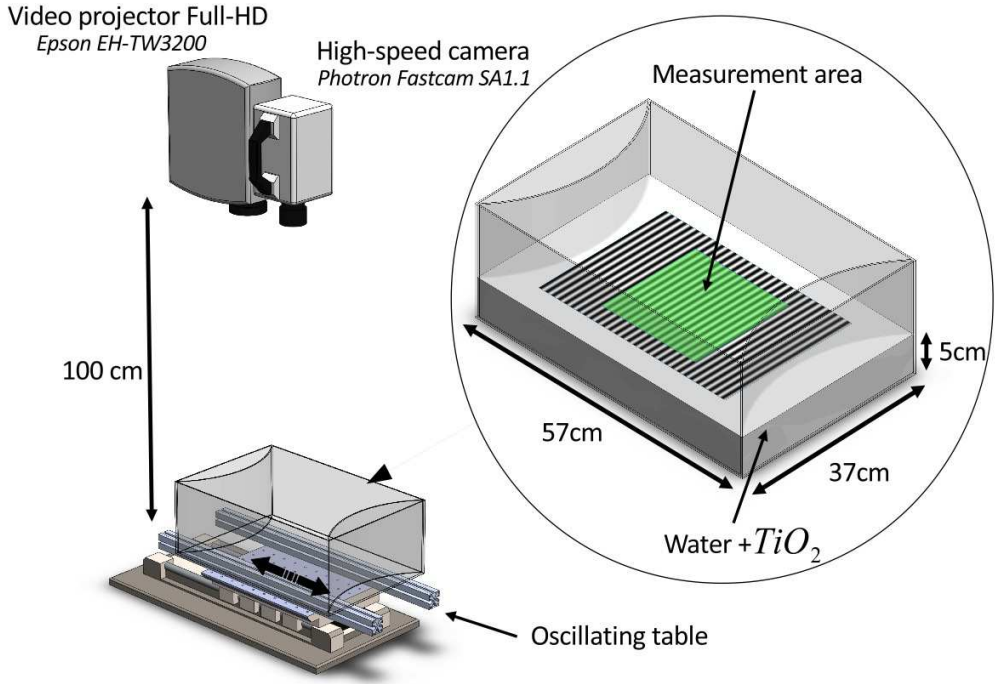}
\caption{Sketch of the experimental setup : Waves are generated by horizontal oscillation of the vessel. Two convex walls are used to improve the isotropy by divergent waves reflexion. The videoprojector creates a pattern at the surface which deformation is recorded by a high-speed camera. The pattern can then be inverted to recover the deformation ~\cite{Cobelli1}. The measurement region is centered and covers a surface $20\times20$~cm$^2$ }
\label{fig1}
\end{figure}
A sketch of the global configuration is shown in figure~\ref{fig1}. The experimental setup is made of a $57\times37$~cm$^2$ wavetank filled with water at a rest height $h_0 = 5$~cm. Surface waves are generated by an oscillating motion of the wave-tank, achieved with an oscillating table. We drive the table with a constant amplitude oscillation which is frequency modulated. The frequency explores the interval [0.5,1.5]~Hz which includes an eigenmode of the wave tank. Two convex walls are used to break the anisotropy of the forcing by divergent waves reflexions. To perform a full measurement of waves in space and time, we use the Fourier transform profilometry ~\cite{Cobelli1}. A videoprojector projects a pattern on the water surface and a high speed camera with a parallel optical axis is focussed on the surface. For the image to form at the very surface, water must diffuse light. As suggested by Przadka {\it et al.}~\cite{Przadka2}, we use anatase titanium dioxide particles (Kronos 1001). These particles do not cause alteration of the pure water surface tension. A concentration of about $40$g/l permit to project a pattern at the water surface with a high contrast level. When waves are present, the pattern is deformed and recorded by the high speed camera. The deformation of the pattern can be then inverted back to the deformation of the surface $\eta(x,y)$. Our measurement area covers a surface of $20\times20$~cm$^2$ at the center of the tank, with $1024\times1024$ pixels resolution at $250$ frames/s. Datasets are composed of 15 movies of $87$~s each to get a good statistical convergence. In this way we obtain measurements of the surface deformation $\eta(x,y,t)$ that are resolved in time and space. We also study the normal velocity field $v(x,y,t)=\frac{\partial \eta}{\partial t}$. Fourier analysis can then be performed by applying a digital Fourier transform in time and/or space.

\begin{figure}[!htb]
\includegraphics[width=8cm]{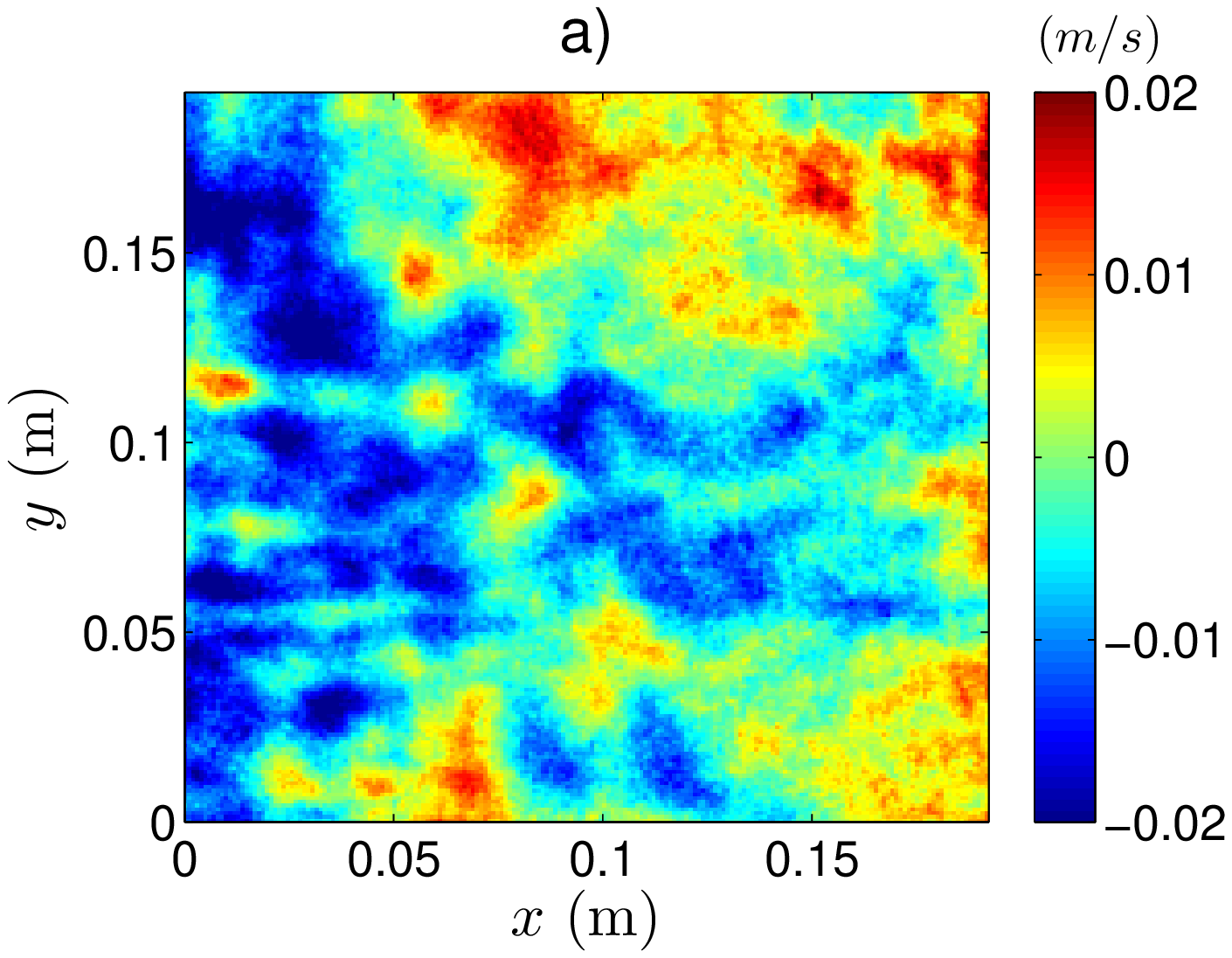}
\includegraphics[width=8cm]{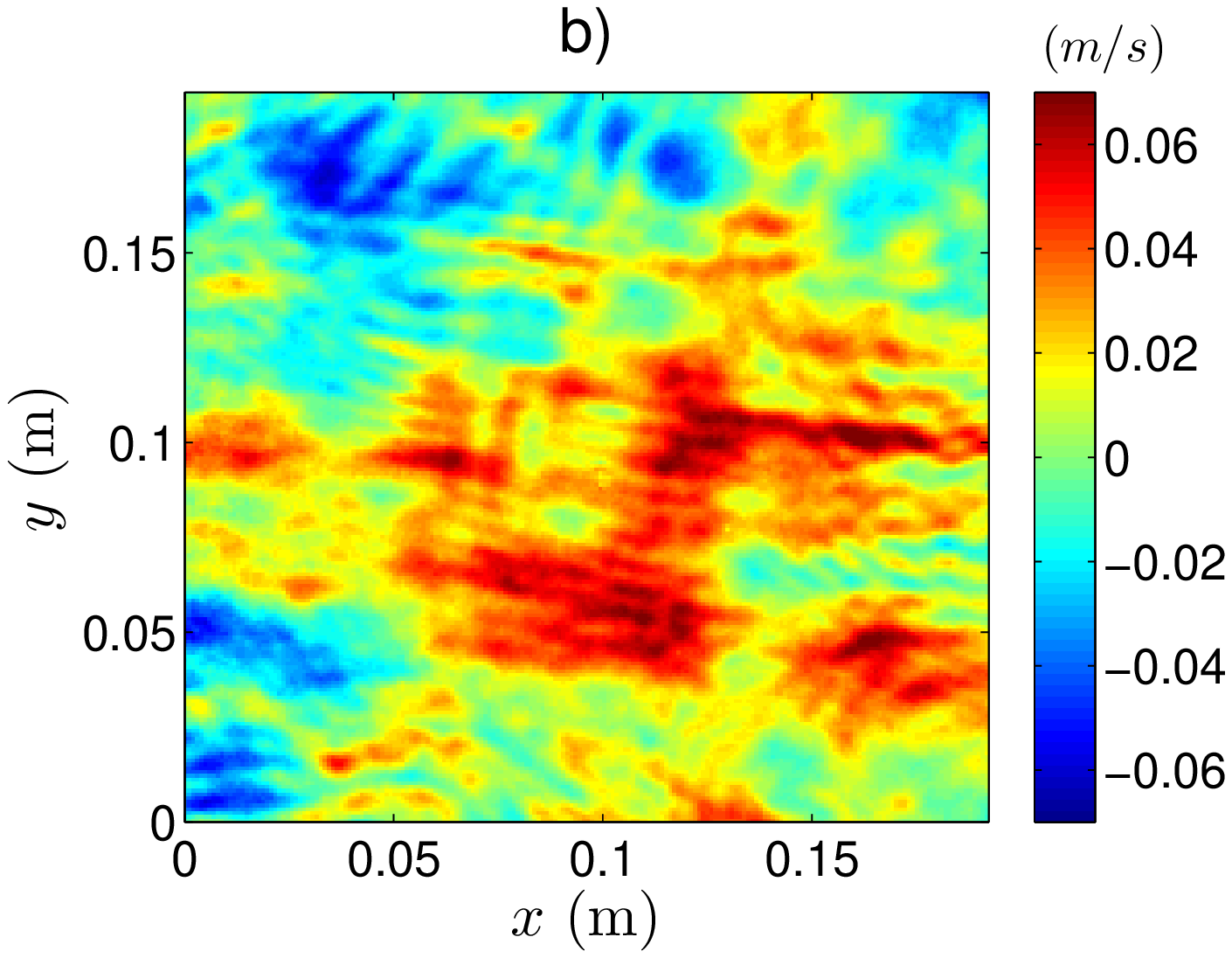}
\caption{(a) vertical velocity field for an experiment for a weak forcing where the typical waves steepness is $\sigma=0.025$. (b)vertical velocity field for a stronger forcing where $\sigma=0.08$.} 
\label{fig2}
\end{figure}
We show in fig \ref{fig2} two snapshots of the vertical velocity field $v=\frac{\partial \eta}{\partial t}$ where $\eta(x,y,t)$ is the altitude of the water surface: (a) weak forcing (b) stronger forcing. To get an estimation of the intensity of the forcing, we define the typical wave steepness $\sigma$:
\begin{equation}
\sigma=\left\langle \sqrt{\frac{1}{S}\int_S \left\|\nabla h(x,y,t)\right\|^2dxdy} \right\rangle
\label{equ2}
\end{equation}
We observe that the stronger forcing is more organized with the presence of coherent waves packets. 
Our experiments range roughly from $\sigma = 0.02$ to $\sigma = 0.1$. The two presented regimes recover the two regimes of water turbulence described by Cobelli {\it et al} ~\cite{CoPr11}. At weak steepness, a regime of truly weakly nonlinear wave turbulence is observed as will be confirmed below. At larger steepness a regime of strongly non linear waves is observed. Part of the strongly nonlinear structures that are observed could actually come directly from the forcing or interactions with the wall. Indeed at very weak forcing, the amplitude of the waves is very small (a couple millimeters) so that the perturbations induced by the depinning of the water contact line on the asperities of the walls are no longer fully negligible. Strong wave amplitude can be generated as well by focussing in the corners of the box. At very strong forcing the velocity of the walls can become close to the celerity of the waves so that the waves that are generated can be strongly nonlinear at the generation. Thus separating the origin of the non linearity between bulk and walls is actually challenging although one expects the bulk effects to dominate when the system is large enough.


\section{Spectral Analysis}

\begin{figure}[!htbp]
\includegraphics[width=16cm]{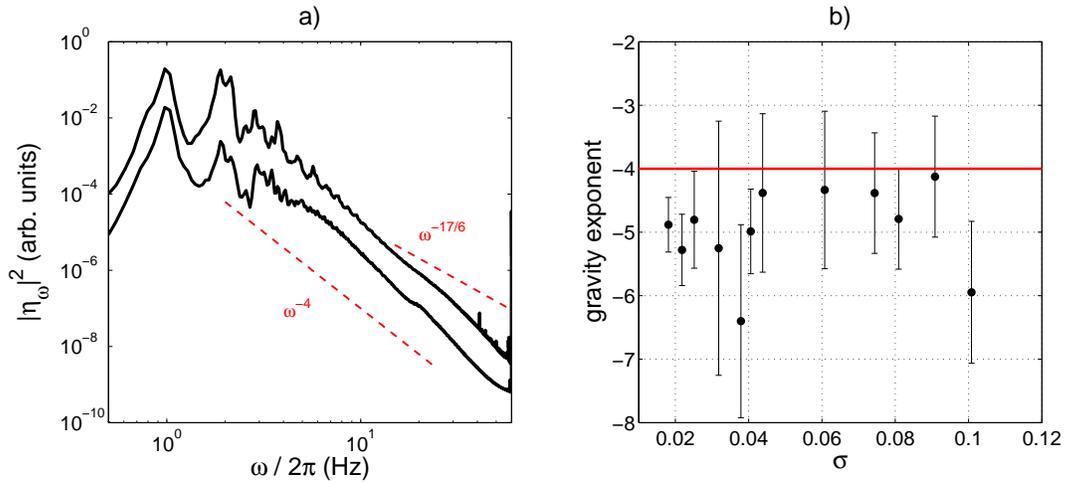}
\caption{(a)Temporal spectrum $E^\eta(\omega) = \langle |\eta(x,y,\omega)|^2 \rangle $ for two experiments where the typical waves steepness is $\sigma=0.025$ (bottom) and $\sigma=0.08$ (top). Dashed red lines represent theoretical slopes predicted by the WTT for pure gravity waves ($\propto \omega^{-4}$) expected at low frequencies and pure capillary waves expected at large frequencies ($\propto \omega^{-17/6}$). (b) Spectral exponent fitted for a set of experiments on the low frequency part of our spectrum.}
\label{fig3}
\end{figure}
In this part we operate a first analysis of the turbulence regime by using Fourier analysis.
By computing the Fourier transform in time of $v(x,y,t)$ we obtain the temporal spectrum $E^\eta(\omega) = \langle |\eta(x,y,\omega)|^2 \rangle $  where $\langle ...\rangle$ denote the average over time windows and space. We compute the time Fourier transform using the Welch method with a $16$~s time window of the signal. Figure \ref{fig3}(a) shows $E^\eta(\omega)$ for the two experiments where $\sigma=0.025$ and $\sigma=0.08$. The forcing operates in the range [0.5,1.5]~Hz and we observe an energy cascade up to $60$~Hz. The theoretical slopes for gravity waves ($\le 15$~Hz) and capillary waves predicted by the WTT are shown in red. We observe a spectrum  much steeper than the theory in the capillary range. This is possibly due to dissipation which becomes significant at these frequencies. Figure~\ref{fig3}(b) shows the exponent extracted by fitting a power law to the temporal spectrum in the gravity range. For a large range of experiments we found an exponent near $-5$ with a significant scattering. Although it is in contradiction with the theoretical prediction, it is actually in agreement with previous observations (~\cite{R10,CoPr11,Issenmann2013,Deik2015}). The scattering may be caused by the pollution of the water surface which generate strong dissipation (by resonance between gravity-capillary waves and Marangoni waves ~\cite{Przadka2}) although experiments have been done with special care : filtered water, careful cleaning and protection of the wavetank to limit the pollution by particles in the air. However, it remains difficult to maintain control of the surface quality over long times.

\begin{figure}[!htbp]
\includegraphics[width=8cm]{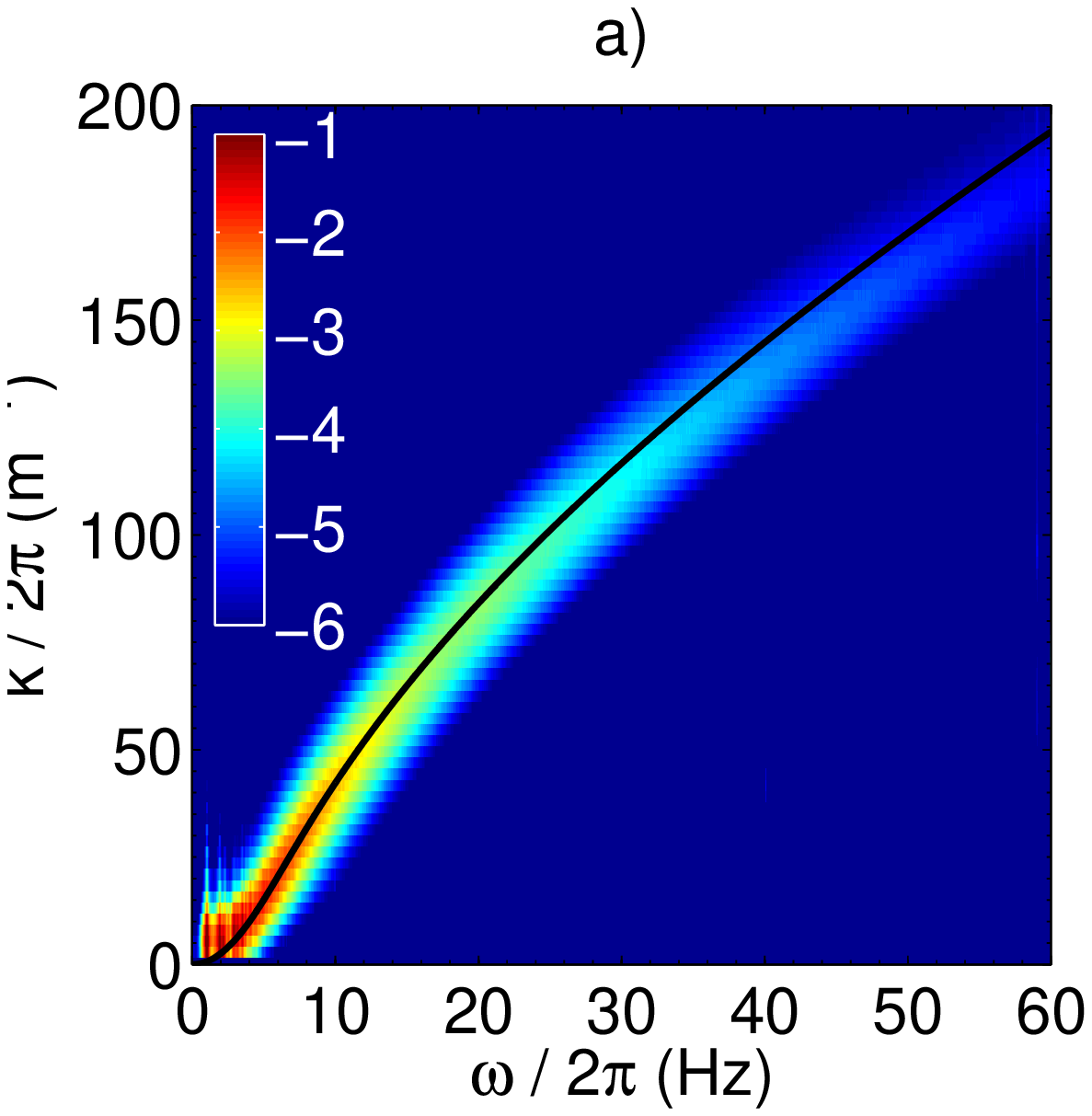}
\includegraphics[width=8cm]{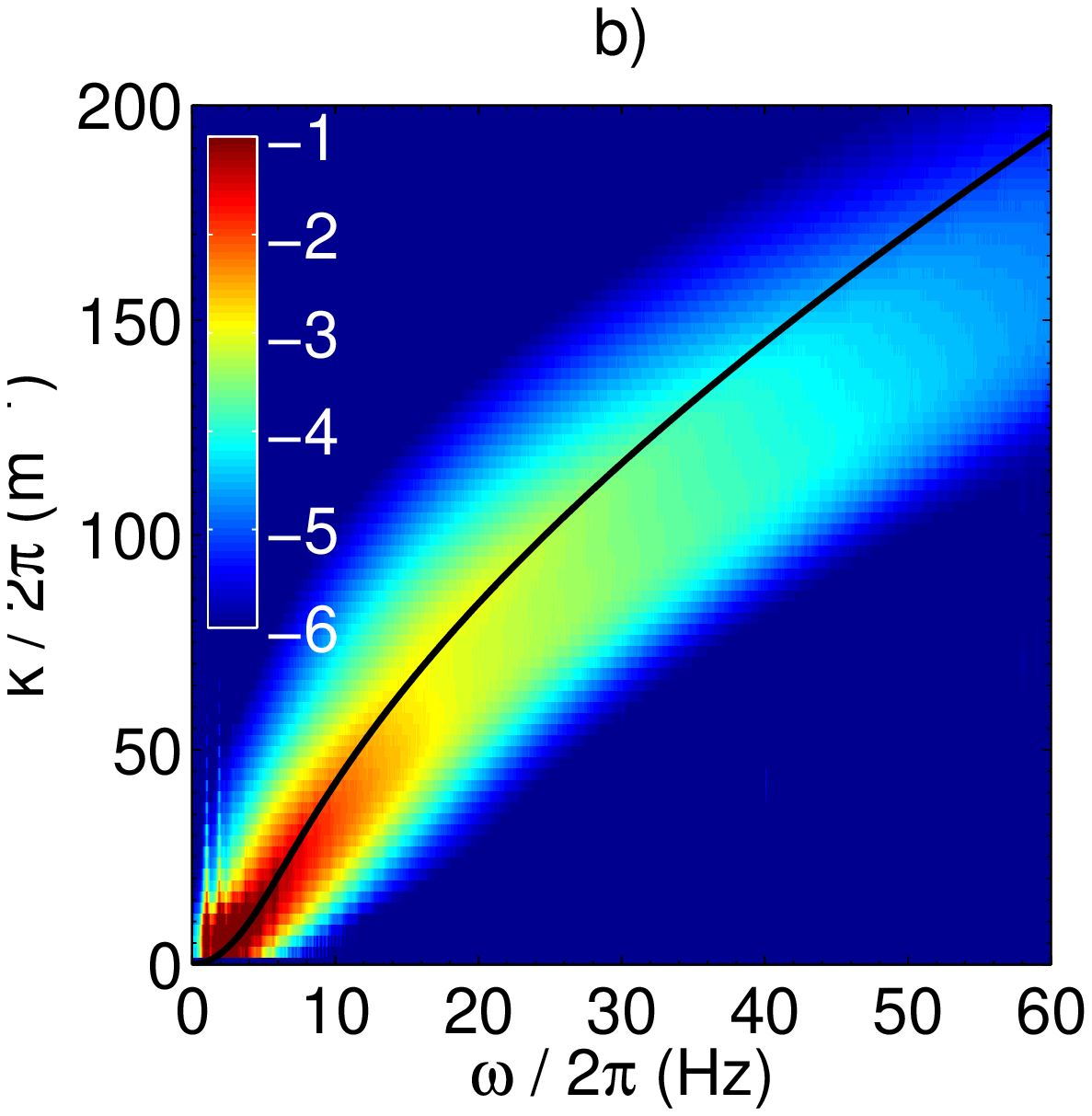}
\includegraphics[width=8cm]{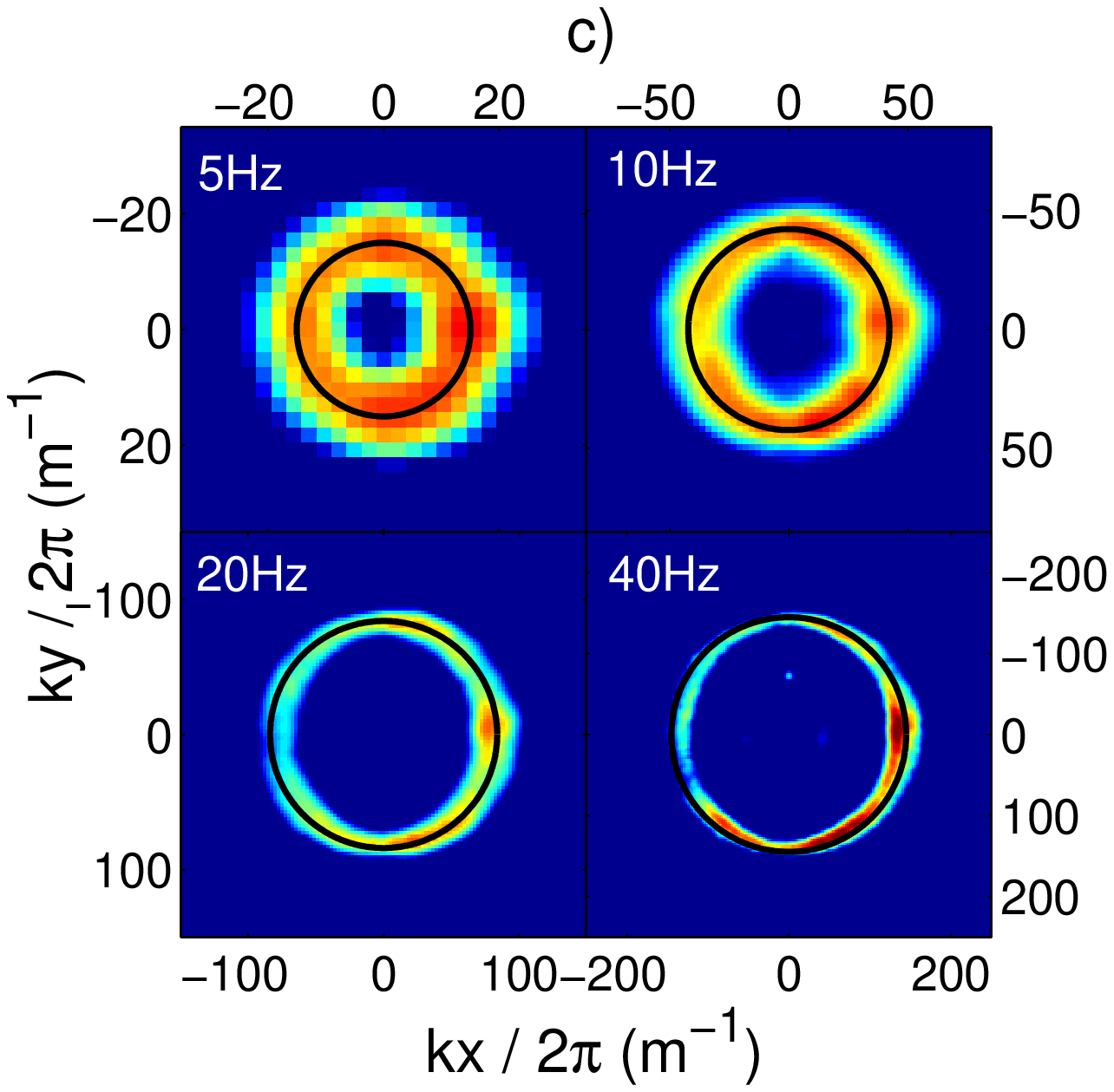}
\includegraphics[width=8cm]{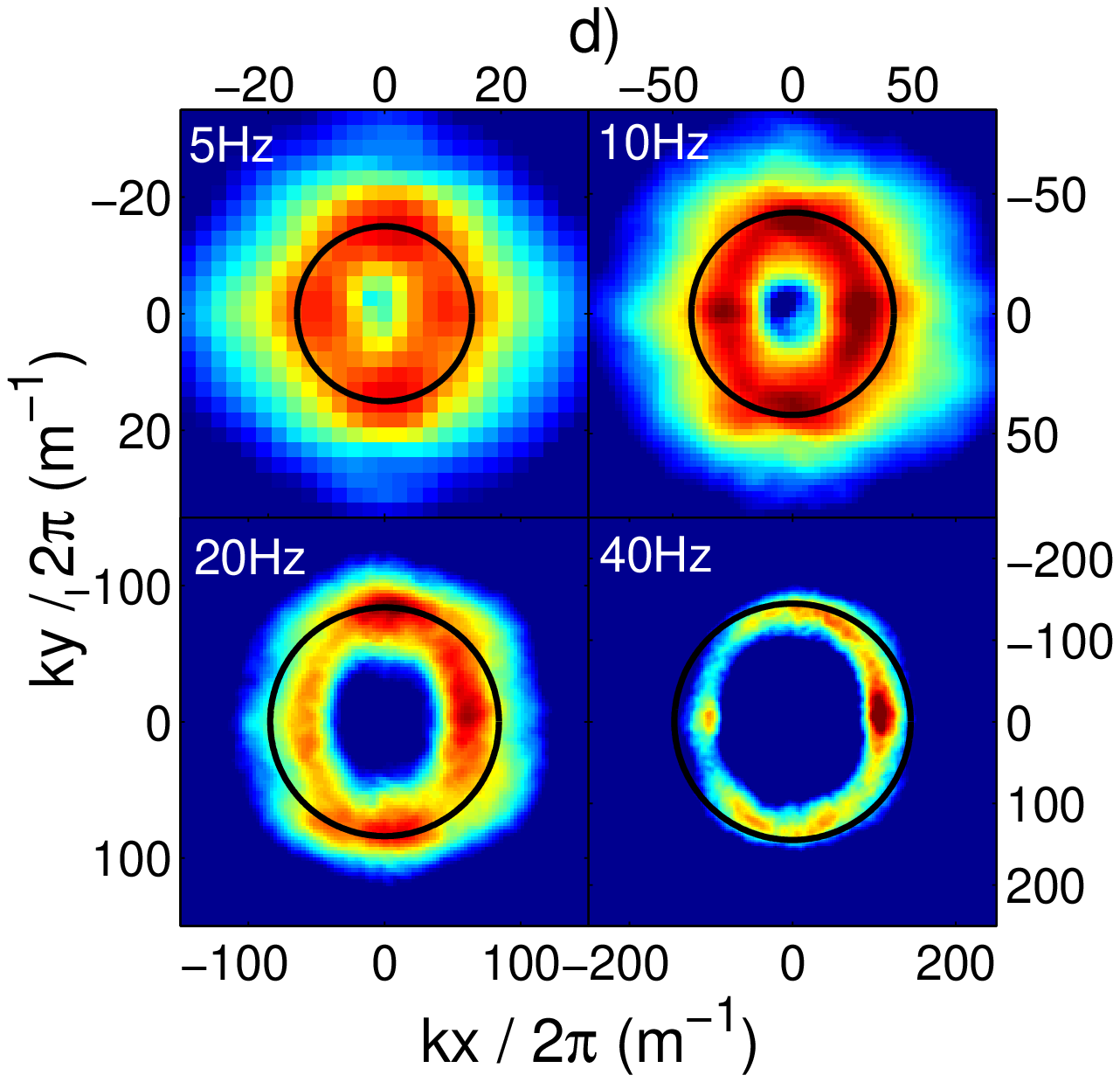}
\caption{(a)Space-time Fourier spectrum of the velocity field of the waves $E^v(k,\omega)$ for an experiment where $\sigma=0.025$. (b) $E^v(k,\omega)$ for an experiment where $\sigma=0.08$. In both cases, the solid black line is the theoretical deep water linear dispersion relation for pure water $\omega^2=gk+\frac{\gamma}{\rho}k^3$ with $\gamma=72$~mN/m. The crossover between gravity and capillary waves occurs at $k_c=\sqrt{\rho g/\gamma}=120 \pi$ corresponding to a wavelength of 1.7~cm and a frequency equal to 13~Hz. (c) cross sections of $E^v(\mathbf k,\omega)$ at given frequencies at weak forcing ($\sigma=0.025$, same as (a)). The solid black line is the linear dispersion relation. (d) same as (c) for the case of stronger forcing.}
\label{fig4}
\end{figure}
The full space-time measurements obtained with the Fourier transform profilometry allow us to compute the complete spectrum $E^v(\mathbf k,\omega)$. The computation of such space-time spectra is demanding in terms of data (notably for 3D systems) but is the adequate tool to discriminate among various sorts of waves or non-propagating motion~\cite{epjb,Herbert,Leoni2,Leoni,Ardhuin}. The spectrum is computed as:
\begin{equation}
E^v(\mathbf k,\omega)=\langle |v(\mathbf k,\omega)|^2\rangle
\label{equ3}
\end{equation}
As for $E^\eta(\omega)$, the signal is divided into $16$~s windows to compute the Fourier transform in time. In order to visualize $E^v( \mathbf k,\omega)$ on a 2d picture, we perform an angular integration of $\mathbf k$ to obtain $E^v(k,\omega)$. Figure~\ref{fig4}(a) shows $E^v( k,\omega)$ at weak forcing. The energy is concentrated around the linear dispersion relation of gravity-capillary waves (\ref{rd}) (black line).
The energy spread observed around the dispersion relation is caused by weak non linear effects: the coherence of the linear waves is affected by the slow nonlinear energy exchanges~\cite{epjb, Miquel3} (note that an intrinsic finite spectral width exists also due to the finite spatial size of the images and to dissipation). The change of the width of the dispersion relation is thus a measure of the degree of non linearity of the system. Figure \ref{fig4}(c) shows four sections of the complete spectrum  $E^v(\mathbf k,\omega)$ at given frequencies. We observe energy in almost all directions with a slight shift from the linear dispersion relation at high frequencies. This weak forcing corresponds to the second regime of turbulence reported by Cobelli {\it et al.} ~\cite{CoPr11}. At increased forcing intensity (fig. \ref{fig4}(b) \& (d)), the spread of the energy around the dispersion relation is much wider, suggesting much stronger non-linearities. 
Furthermore, a non-linear shift is observable for all the frequencies. This non-linear shift appears to be much stronger in the direction of the forcing ($Ox$) than in the transverse direction ($Oy$). This is visible in the cross sections of the spectrum in fig.~\ref{fig4}(d) that are no longer circular as in fig.~\ref{fig4}(c). This deformation of the spectrum seems to be related with the amplitude of the forcing and is also qualitatively consistent with observations by Berhanu \& Falcon~\cite{Berhanu2013} who interpret this shift as being due to nonlinear corrections due to large amplitude low frequency modes. They suggest that the dispersion relation could be 
\begin{equation}
\omega^2=\left(gk[1+(ak)^2]+\frac{\gamma}{\rho}k^3\left[1+\left(\frac{ak}{4}\right)^2\right]^{-1/4}\right)\tanh(kh)
\end{equation}
where $a$ is the {\it rms} wave amplitude. If this effect is actually depending on the propagation direction of the low frequency Fourier modes, it could be responsible for the shift in the direction of the forcing for which the low frequency components are stronger.

Thus our observations confirm the existence of two distinct regimes (weak and strong) and the following part is devoted to a detailed study of 3-wave interaction that is supposed to be at the core of the energy cascade.

\section{3-wave resonances}

The most natural way to investigate non linear coupling among $N$ waves is the use of $N$-order correlations such as
\begin{equation}
\langle a_1a_2...a_pa_{p+1}^*...a_N^*\rangle
\end{equation}
where $a_i$ is the wave amplitude in frequency space $a_i(\omega_i)$ or in wavevector space $a_i(\mathbf k_i)$ ($.^*$ is the complex conjugation). This sort of generic correlations probes resonances $\omega_1+\omega_2+...+\omega_p=\omega_{p+1}+...+\omega_{N}$ or $\mathbf k_1+\mathbf k_2+...+\mathbf k_p=\mathbf k_{p+1}+....+\mathbf k_N$ respectively. In the following we will investigate the 3-wave case $N=3$ both in frequency and wavenumber space in order to get detailed information on the geometry of the wave interactions.

\subsubsection{Theoretical analysis}
\label{theo}

\begin{figure}[!htb]
\includegraphics[width=10cm]{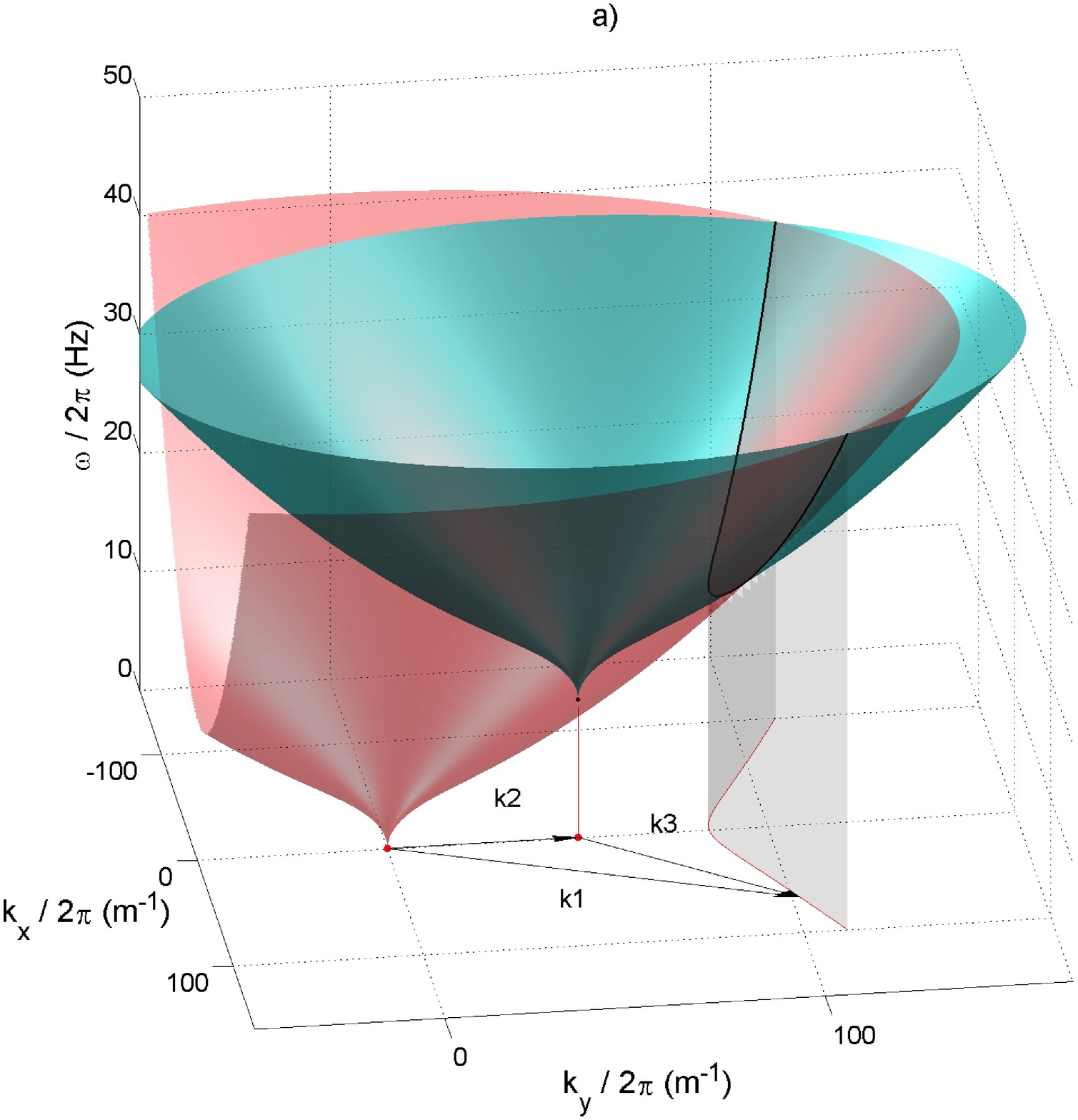}
\includegraphics[width=12cm]{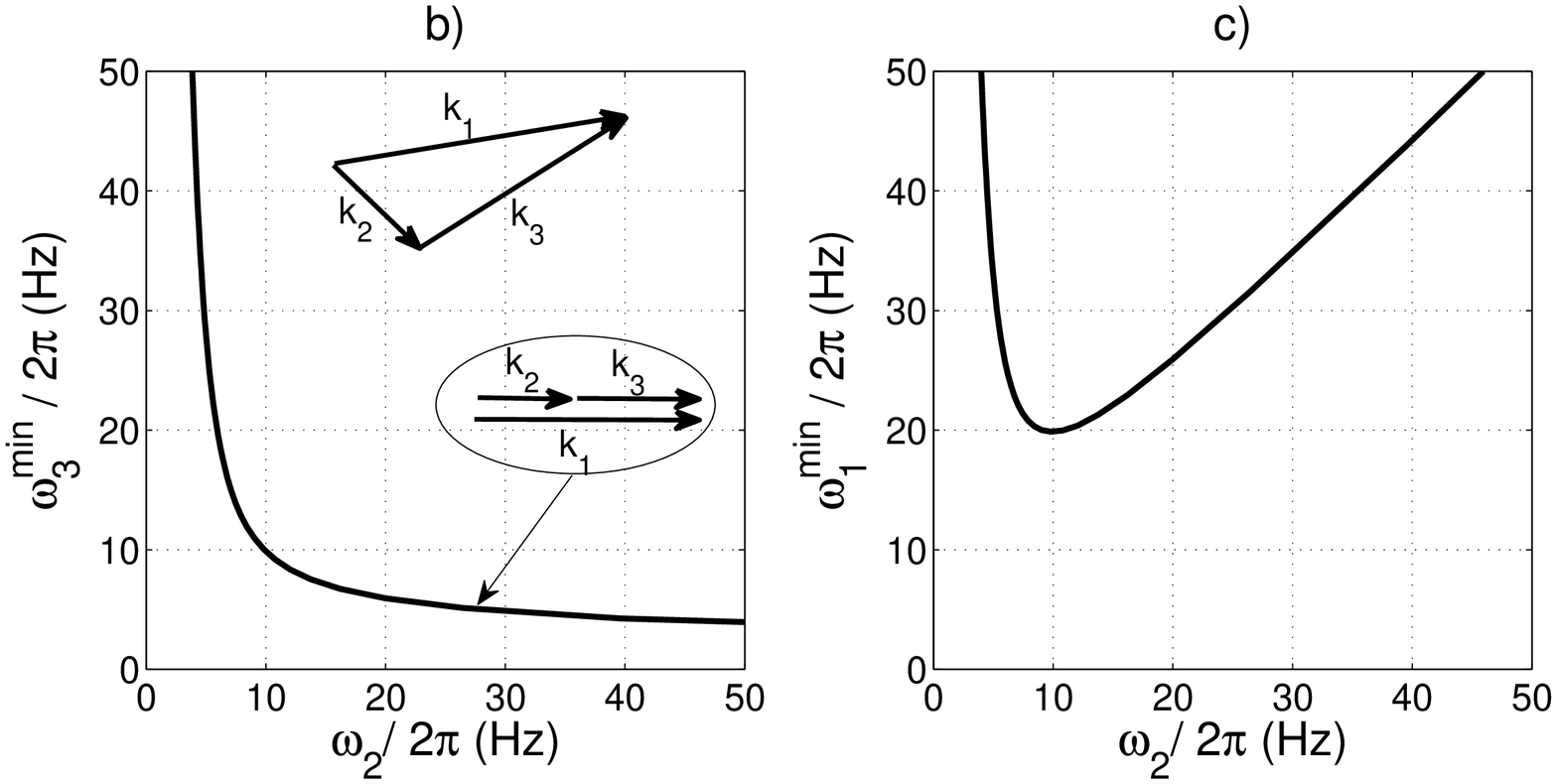}
\caption{3-waves resonant solutions for gravity capillary waves. (a) Representation in the 3D-space. The red surface is the linear dispersion relation and a first wave [$\mathbf k_2$,$\omega_2$] is given. A second dispersion relation is then plotted from this point (green surface). Thus, the resonant solution, satisfying the equations (\ref{equ4}) and (\ref{equ5}), are the intersection between this two surfaces (thick black line). A projection of the solutions in the $[k_x,k_y]$ space is plotted in red, showing then the possible wavenumbers $[\mathbf k_1,\mathbf k_2,\mathbf k_3]$ combinations. (b) $\omega_3^{min}$ as a function of $\omega_2$. $\omega_3^{min}$ is the minimum value of $\omega_3$ for which resonant interaction are possible for a given $\omega_2$. (c) corresponding value $\omega_1^{min} = \omega_3^{min} + \omega_2$.}
\label{fig5}
\end{figure}
The order of non linear wave interaction depends on the order of the non linearity and the possibility or not to have solutions for the resonance equations. McGoldrick and then Simmons ~\cite{Mcgoldrick1965,Simmons1969} have investigated the 3-waves resonant solutions in gravity-capillary regime satisfying these two equations

\begin{equation}
\mathbf{k_1}=\mathbf{k_2}+\mathbf{k_3}
\label{equ4}
\end{equation}
\begin{equation}
\omega_1=\omega_2+\omega_3
\label{equ5}
\end{equation}

Figure \ref{fig5}(a) displays a full space-time representation of the resonant solutions for a given wave $[\mathbf k_2,\omega_2]$. The red surface shows the linear dispersion relation (\ref{rd}) and marks the point $[\mathbf k_1,\omega_1]$. As $\omega_1=\omega_2+\omega_3$, the green surface displays the dispersion relation shifted to start from the point $[\mathbf k_2,\omega_2]$ and thus marks the points $[\mathbf k_2+\mathbf k_3,\omega_2+\omega_3]$. Thus the resonant manifold is the intersection of the two surfaces marked as the thick black line. Due to the change of curvature of the dispersion relation, the intersection exists whatever the chosen value for $\omega_2$.

The resonant solutions show a point of minimum distance to the origin denoted [$\mathbf k_3^{min}$,$\omega_3^{min}$] which correspond to the special 1D-configuration where the three wave vectors $\mathbf k_1$, $\mathbf k_2$ and $\mathbf k_3$ are collinear. Figure~\ref{fig5} (b) displays the variation of $\omega_3^{min}$ as a function of $\omega_2$. In the whole area above this line exact resonant interaction are permitted. Thus, we observe the possibility of strong non local coupling: a low frequency gravity wave can interact only with capillary waves of much higher frequency. As it is well known, we see that pure gravity waves ($\omega /2\pi < 10$~Hz) can not interact with each other through 3-waves interaction.  Fig.\ref{fig5} (c) shows $\omega_1^{min} = \omega_3^{min} + \omega_2$ as a function of $\omega_2$  to enhance the presence of a overall minimum of $\omega_1$. This special point correspond to the degenerate case of Wilton waves : $\omega_1=2\omega_2=2\omega_3=2\pi\times19.6$~Hz (fig.~\ref{fig2}(c)) ~\cite{wilton1915lxxii}. The presence of such 3-waves interactions have been verified experimentally by Hammack {\it et al.} ~\cite{Henderson1987}

\begin{figure}[htb]
\includegraphics[width=16cm]{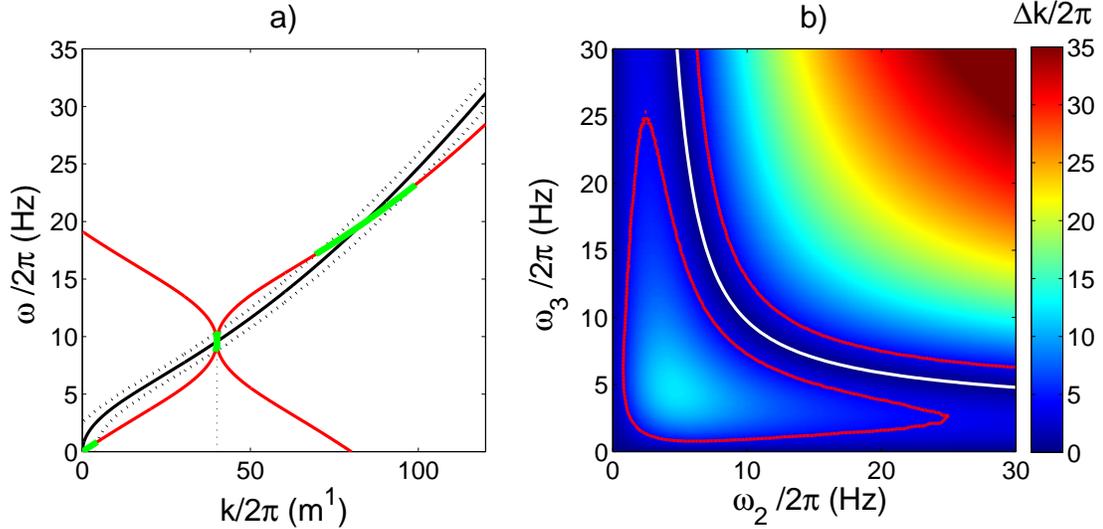}
\caption{(a) approximate resonant solution in the 1D case for a chosen $k_2$ (see text). The dotted black lines represent a non-linear widening of the dispersion relation by $\delta k /2\pi = 4$~m$^{-1}$ (which order of magnitude is estimated  from $E^v(\mathbf k,\omega)$ ). The green lines are the approximate solutions. (b) Color map of the distance $\Delta k$ between the two dispersion relation (the black and the red lines plotted in (a)) the in the frequency space. The white line is the exact solution corresponding to $\Delta k /2\pi = 0$~m$^{-1}$. The red line represent $\delta k /2\pi= 4$~m$^{-1}$ which bound the domain where approximate resonance occur.}
\label{fig6}
\end{figure}
As shown above in fig.~\ref{fig4}, non-linear effects generate an energy spread around the dispersion relation. This may be seen as some uncertainty on the dispersion relation. We want to investigate the effect of this uncertainty on the resonances. For simplicity we consider only a unidirectional configuration of the wave vectors and we attribute all the (constant) uncertainty $\delta k$ to the wave vector equation. In this simplified case, equation (\ref{equ4}) can be rewritten as:

\begin{equation}
k_1 \pm {\delta k} =k_2+ k_3
\label{equ6}
\end{equation}
\begin{equation}
\omega_1=\omega_2+\omega_3
\label{equ7}
\end{equation}

Figure \ref{fig6} (a) shows the solution for a given wave $k_2 /2\pi = 40$~m$^{-1}$. The exact solution similar as in fig.~\ref{fig5} correspond to the intersection between the black and the red curve. The non-linear widening $\delta k /2\pi = 4$~m$^{-1}$ (order of magnitude estimated from $E^v(\mathbf k,\omega)$ of the weak forcing in fig.\ref{fig4}) is represented by the two dotted black lines near the thick black line of the dispersion relation. The range of new solutions permitted by this non-linear widening are represented by the green lines. Figure~\ref{fig6}(b) shows the distance $\Delta k$ between the black and the red lines as a function of frequency pairs $(\omega_2,\omega_3)$. The white line in fig.~\ref{fig6}(b) is the case of exact resonances for which the distance is zero. The thin red line limits the area in which the distance $\Delta k$ is smaller than $\delta k$ in which approximate resonances are possible. The range of possible resonances is thus strongly increased and thus these approximate resonances can increase significantly the intensity of energy transfer. In particular in a bounded system, where the discretization may not allow exact solutions, interaction may exist nonetheless. Note that approximate resonances also allows 3-waves interaction in the gravity regime whereas exact resonances are only possible among 4-waves. Indeed, as the two dispersions relation are very close for these gravity waves, a weak non-linear widening permits 3-waves interactions. This analysis could be extended to the 2d case (fig.\ref{fig5}) and we could also make $\delta k$ vary as a function of $\omega$ to be closer of the experiments but in both cases we do not expect a significant change to the simple 1D case.

\subsubsection{Frequency correlations}

\begin{figure}[!htb]
\includegraphics[width=16cm]{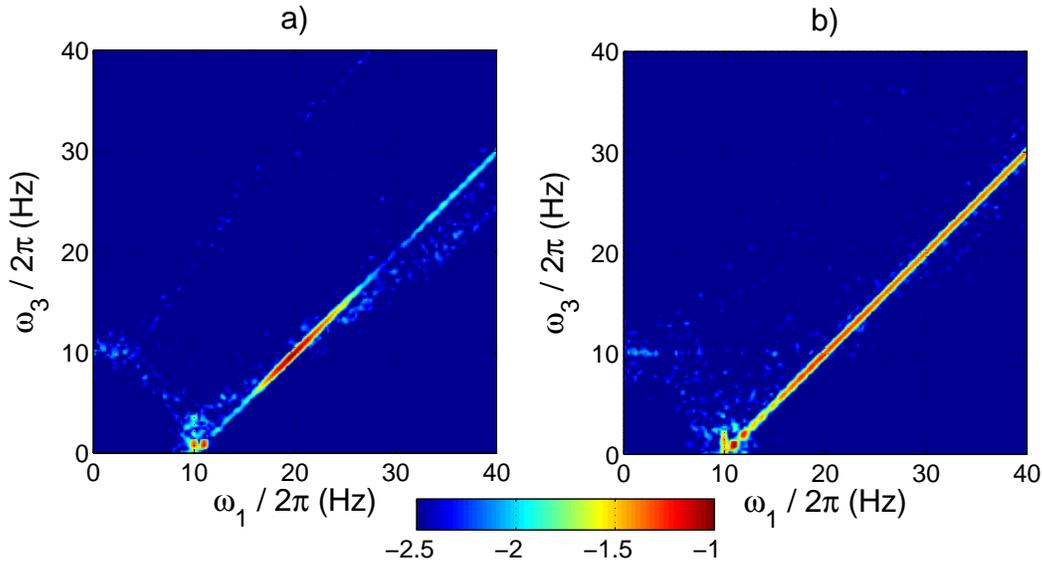}
\caption{Representation of 3-wave coherence $C_\omega(\omega_1,\omega_2,\omega_3)$ where $\omega_2 /2\pi =10$~Hz (see text for definitions): (a) for the weak forcing ($\sigma=0.025$) (b) for a strong forcing ($\sigma=0.08$). The surrounding blue background corresponds to statistical convergence noise so that the statistical convergence is about $10^{-2.5}$. The observed line of high correlation is above the convergence level by more than one order of magnitude at the maximum showing the presence of significant 3-wave coupling in the signal. The line remains on the resonance curve $\omega_1 = \omega_2+ \omega_3$ as expected from such a statistical estimator for a stationary random signal.  Correlations are $log_{10}$ color-coded.}
\label{fig7}
\end{figure}
As we expect 3-wave resonances, we focus our study on 3-wave correlations. The simplest case is to work in the frequency space due to the reduced number of parameters: three frequencies reduced to two using the resonant equation (\ref{equ5}). From $v(x,y,t)$, we compute the Fourier transform in time over $4$~s time windows so that to obtain $v(x,y,\omega)$. Correlations are then computed as
\begin{equation}
C_\omega(\omega_1,\omega_2,\omega_3)=\frac{|\langle\langle v^\star(x,y,\omega_1) v(x,y,\omega_2)v(x,y,\omega_3)\rangle\rangle |}
{\left[E^v(\omega_1)E^v(\omega_2)E^v(\omega_3)\right]^{1/2}}
\label{equ8}
\end{equation}
where $\langle\langle...\rangle\rangle$ denotes a double average: a space average over $(x,y)$ on the image and over successive time windows. $E^v(\omega) = \langle\langle |v(x,y,\omega)|^2 \rangle\rangle $ is the frequency spectrum. The coherence defined in this way should be 0 for uncorrelated waves and 1 for perfectly correlated waves.  For a stationary random signal such this correlation is expected to be non zero only along the resonance line $\omega_1 = \omega_2 + \omega_3$. Figure~\ref{fig7} shows the coherence map at a given frequency $\omega_2 /2\pi =10$~Hz. We observe for both forcing intensities that a line of correlation emerges from the statistical convergence noise which is less than $3.10^{-3}$. This implies the presence of a significant 3-waves resonant process in the signal.  We note a difference between the two forcing intensities: at weak forcing a correlation peak is seen near 20~Hz whereas the strong forcing appears much more homogeneous in terms of variation of the correlation amplitude with the frequency. For exact resonances, with a given value of $\omega_2=10$~Hz, the minimum value of $\omega_3$ which is allowed is also 10~Hz and the corresponding minimum value of $\omega_1$ is about 20 Hz (see fig.~\ref{fig5}). Thus at the weakest forcing one expects that the correlation becomes non zero only for values of $\omega_1$ larger that 20~Hz. The fact that the correlation is weak for values of $\omega_1$ larger than 20~Hz suggests that the unidirectional coupling (corresponding to $\omega_1=20$~Hz) is dominant. At larger forcing intensity the approximate resonances become numerous and thus the wave coupling involves many more wave triads.

\begin{figure}[!htb]
\includegraphics[width=16cm]{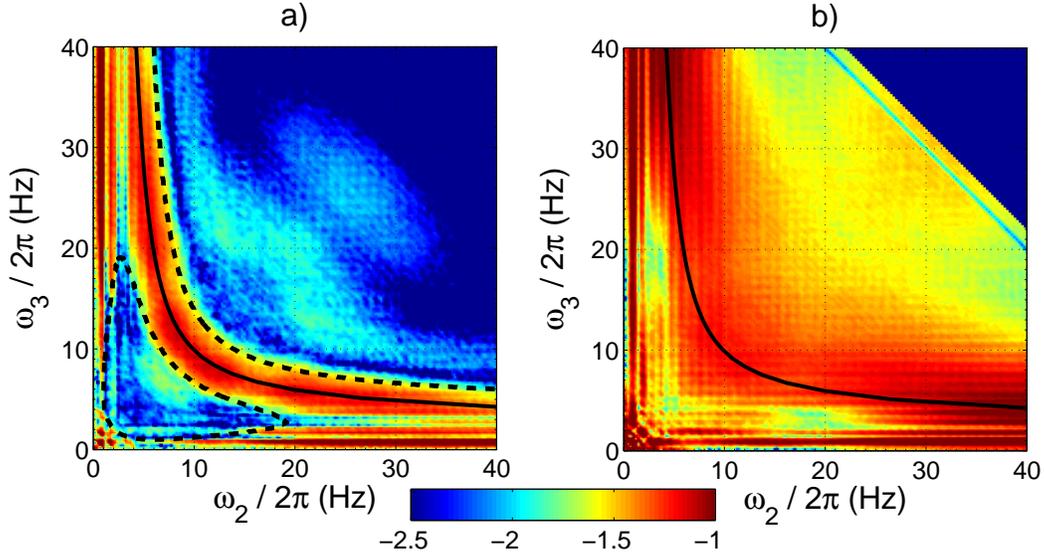}
\caption{Bicoherence $B_\omega(\omega_2,\omega_3)$ for a weak forcing where $\sigma=0.02$ (a) and a strong forcing where $\sigma=0.08$ (b). The solid black line represent the border $\omega_3^{min}$ of the area of possible exact resonant solutions. The dotted black lines represent the 1D solution affected by a non-linear spectral spreading $\delta k /2\pi=4$~m$^{-1}$. Correlations are $log_{10}$ coded}
\label{fig8}
\end{figure}
In order to perform a deeper analysis of the resonant interaction, we compute the bicoherence, defined as 
\begin{equation}
B_\omega(\omega_2,\omega_3)=C(\omega_2+\omega_3,\omega_2,\omega_3)
\label{equ9}
\end{equation}
It corresponds to the extraction of the resonant line observed in the fig.\ref{fig7} for all frequencies $\omega_2$. The presence of this converged line above the noise level have been checked for all frequencies. Figure~\ref{fig8} shows the bicoherence map at the two forcing amplitudes. In the weak forcing case (a), a noticeable organization of the coherence is seen. First the figure is symmetric along the diagonal $\omega_2=\omega_3$ as expected from the symmetry of the resonance relations when permuting the indices $2\leftrightarrow3$. Second a wide line of high coherence (about $10^{-1}$) is located close to the $\omega_3^{min}$ curve (thick black line) described previously and corresponding to exact resonances in the unidirectional case. In the region of allowed exact resonances (above the thick black line) the coherence level above this curve remains low ($\lesssim 10^{-2}$). This suggests that the dominant interactions are close to unidirectionnal. In this case the observed isotropy of the spectrum could be due mostly to the fact that the curved walls create a chaotic cavity. Nevertheless experiments performed in a rectangle cavity do not show much difference in the isotropy. This is most likely due to the fact that the coupling still permits some angular spreading or energy especially at low wavenumber as is confirmed by the wavenumber resonance analysis below.

At very low values of both $\omega_2$ and $\omega_3$ a few spots of high coherence exist that correspond to the allowed approximated 3-wave resonances among gravity waves. 

We observe another region of high coherence for small values of $\omega_3$ and large values of $\omega_2$ (bottom part of fig.~\ref{fig8}(a)). This region is composed of two strong lines of high coherence, corresponding to the frequencies of the first eigenmodes of the wavetank. We can explain the existence of this region with the approximate resonances discussed above. As the interactions are observed to be mostly 1D, we show as the dotted black lines the region of approximate resonances in 1D case for $\delta k /2\pi = 4$~m$^{-1}$ already shown in fig.~\ref{fig6}(b). This solution encircles very convincingly the high correlation area and highlights the importance of these approximate resonances : the number of effective interactions increase significantly and they do operate. The case of stronger forcing is displayed in fig.~\ref{fig8}(b). The coherence is less organized and less straightforward to interpret. The maximum level of correlation is of the same order as the weak forcing but the 1D resonant line is less visible. The spread of the high coherence region is much wider than in the low forcing case. The coherence level is much higher also in the regions that appear in blue in fig.~\ref{fig8}(a)). Wave interactions seem possible in all the domain and there are no longer any indications of a possibly dominant unidirectional process. We also note that some discretization become apparent at higher frequencies. It is consistent with the peaks on the temporal spectrum visible on the fig.\ref{fig3}a. These peaks are most likely related to the strong forcing although their interpretation is not fully clear so far.

\subsubsection{Space analysis}

\begin{figure}[!htb]
\includegraphics[width=15cm]{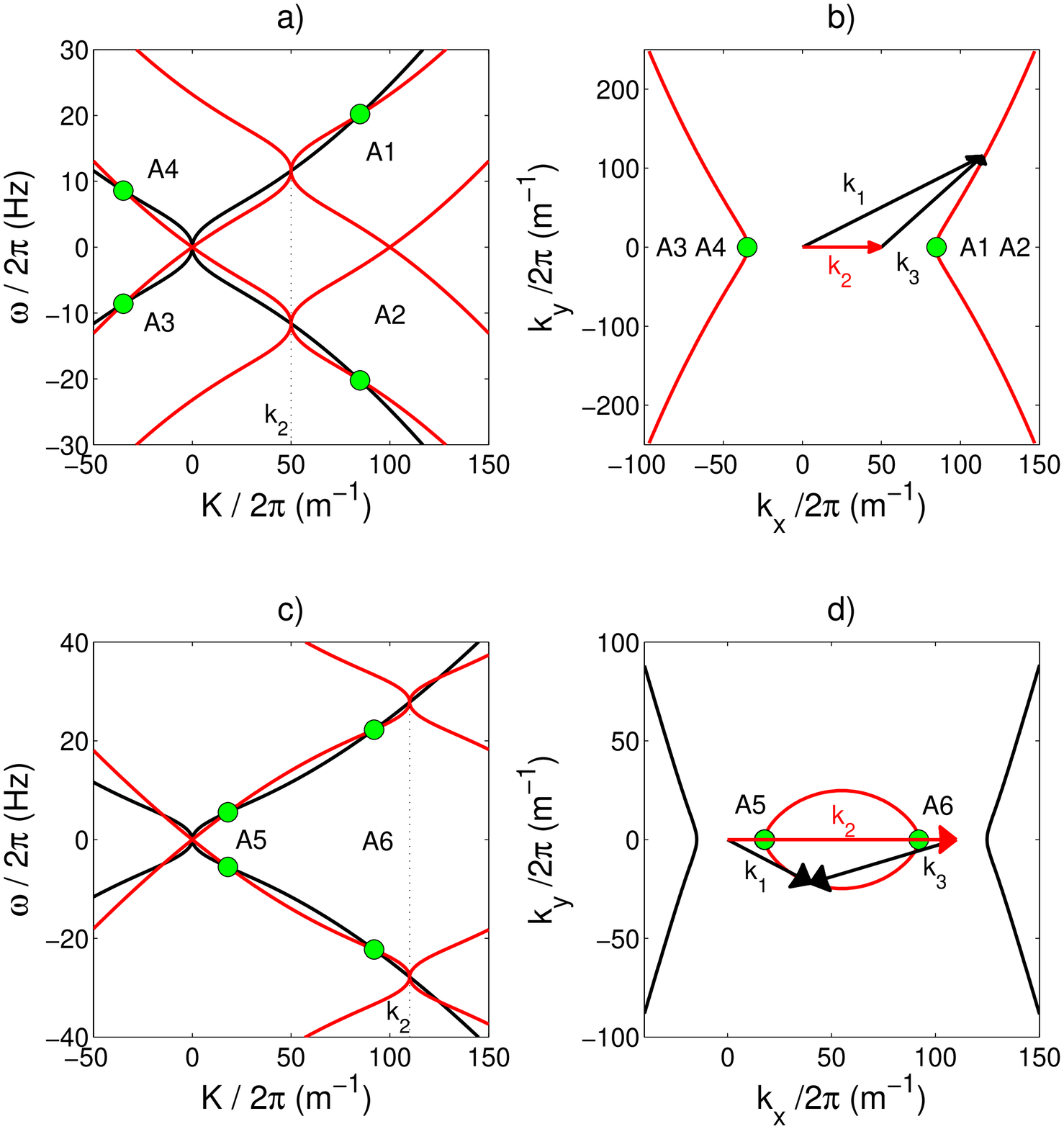}
\caption{(a) Theoretical exact solutions in the unidirectional case for a given wavenumber $k_2 /2\pi=50$~m$^{-1}$. The black line is the dispersion relation and the red lines are the shifted dispersion relations as in paragraph~\ref{theo} and the resonant waves are found at their intersection (green dots). (b) Theoretical exact solutions in the general case, in the $(k_x,k_y)$ space  and for a given wavevector of modulus equal to $k_2/2\pi=50$~m$^{-1}$. Exact 1D solution corresponding to (a) are reported by the green points. Arrows represent a resonant triplet. 
(c) 1D theoretical exact solutions for a larger wavenumber $k_2/2\pi=110$~m$^{-1}$. We observe the apparition of a new region of solutions in addition to the previous observed in (a). This new solution is actually equivalent to the other by permutation of the indices of the waves. (d) 2D theoretical exact solutions in the $(k_x,k_y)$ space for a given $k_2/2\pi=110$~m$^{-1}$. }
\label{fig9}
\end{figure}
In order to confirm directly the unidirectional character of the interactions, we investigate the correlations in the wavenumber domain. Figure~\ref{fig9} shows the expected exact solutions projected in the $[k_x,k_y]$ space. It is a more detailed view of the projections of the solution seen in the fig.~\ref{fig5}(a). Figure~\ref{fig9}(a) illustrate the method to find the resonances: for simplicity we choose the 1D case at a given $\mathbf k_2$ with the inclusion of the negative frequencies and wavenumbers. The black line is the dispersion relation and the red lines are the shifted dispersion relations as in paragraph~\ref{theo}. The resonant solutions are thus the intersections of the black and red curves and they are marked as green dots.

Figure~\ref{fig9}(b) represents the general 2D solutions in the [$k_x,k_y$] space for a given wavevector $\mathbf k_2$. In this representation, exact 1D solutions appear as points and are reported as the green dots. The arrows show an example of allowed wavenumber combination for the given vector $\mathbf k_2$ (in red) . The symmetry in frequency imply that solutions A1 and A2 (as for A3 and A4) are superposed in the [$k_x k_y$] space. To go further, the two branches of solutions are equivalent and correspond to permutations of indices in the resonant equations. For instance the solution A1 correspond to $\omega_1=\omega_2+\omega_3$, and the solution A4 correspond to $\omega_1=-\omega_2+\omega_3$ which can be rewritten as $\omega_3=\omega_1 + \omega_2$. Thus, by permuting the indices, they are fundamentally equivalent. The sub-figures~\ref{fig9}(c) and (d) show the apparition of a new solution at larger $k_2$. As we have just seen previously, they are also equivalent by permutation of the indices.

Similarly to the definition of $C_\omega(\omega_1,\omega_2,\omega_3)$ in the frequency case, we define the correlations in space $C_k(\mathbf k_1,\mathbf k_2,\mathbf k_3)$ as 
\begin{equation}
C_k(\mathbf k_1,\mathbf k_2,\mathbf k_3)=\frac{|\langle v^\star(\mathbf k_1,t) v(\mathbf k_2,t)v(\mathbf k_3,t)\rangle |}
{\left[E^v(\mathbf k_1)E^v(\mathbf k_2)E^v(\mathbf k_3)\right]^{1/2}}\, .
\label{equ10}
\end{equation}
Here the average $\langle...\rangle$ stands for an average over the time. $E^v(\mathbf k) = \langle |v(\mathbf k,t|^2 \rangle $ is the spatial Fourier spectrum.

We thus define the bicoherence $B_k(\mathbf k_2,\mathbf k_3)$ as
\begin{equation}
B_k(\mathbf k_2,\mathbf k_3)=C(\mathbf k_2+\mathbf k_3,\mathbf k_2,\mathbf k_3)
\label{equ11}
\end{equation}

\begin{figure}[!htb]
\includegraphics[width=15cm]{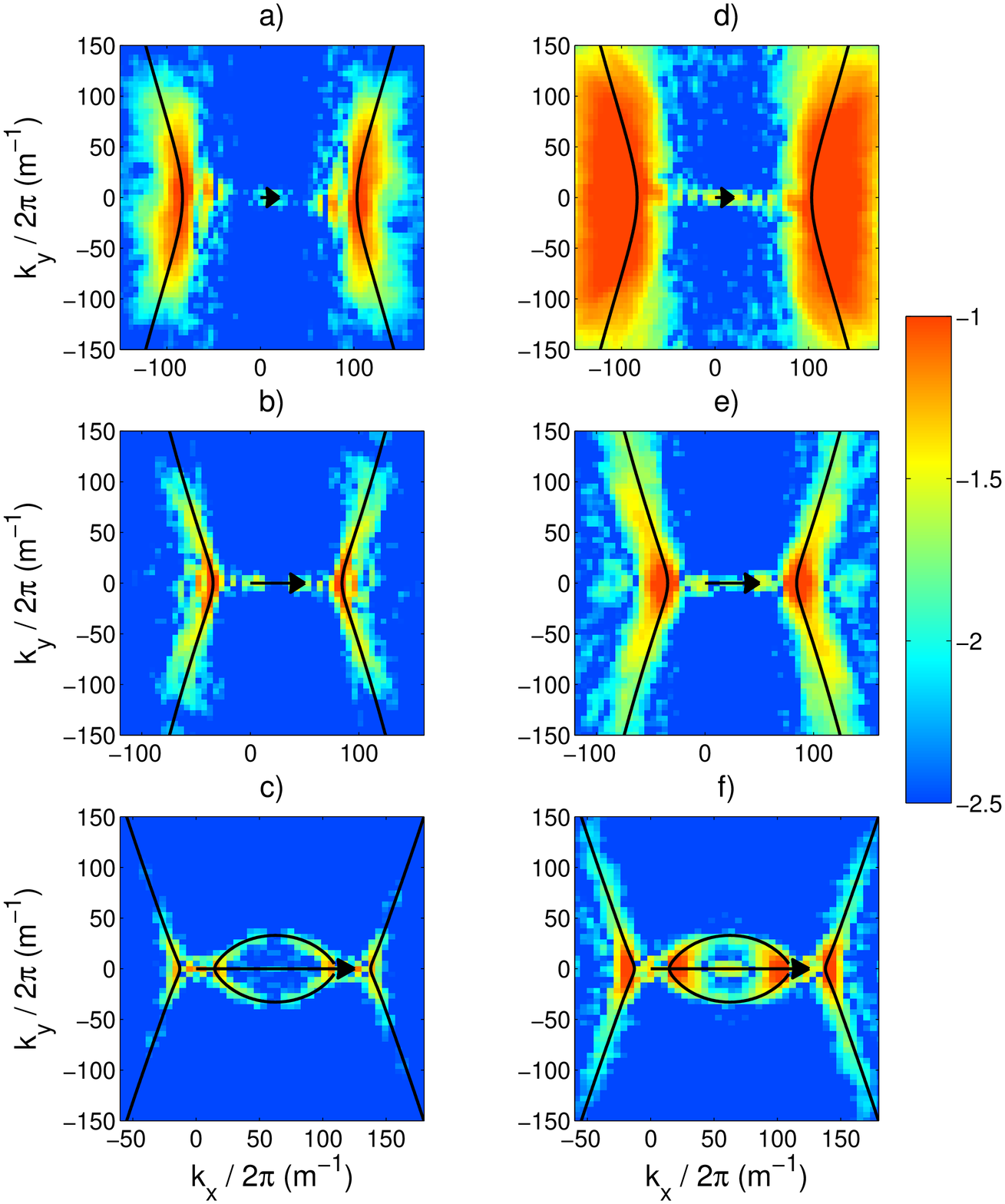}
\caption{Bicoherence $B_k(k_2,\mathbf k_3)$ for $\sigma=0.02$ (left) and $\sigma=0.08$ (right). We fixed $| \mathbf k_2 | /2\pi = 20$ (a,d), $| \mathbf k_2 | /2\pi = 50$ (b,e),  $| \mathbf k_2 | /2\pi = 125$ (c,f). Correlations are averaged over eight different directions for $\mathbf k_2$ (after rotation of the whole picture).The black arrow represent the chosen $\mathbf k_2$ and exact solution are plotted in black. Color is $log_{10}$ coded. We observe a concentration of correlation emerging from the statistical noise along the exact solutions. The spreading along the line come from the approximate resonances. The presence of maximum along the direction of $\mathbf k_2$ confirm the dominance of the unidirectional interactions. For low $\mathbf k_2$ we note a significant angular spreading of interactions.}
\label{fig10}
\end{figure}
Unlike frequency correlation, $B_k(\mathbf k_2,\mathbf k_3)$ is a function of 4 real parameters which makes its representation more difficult. In order to obtain a 2D picture which is easier to analyze, we first chose a value of $|\mathbf k_2|=k_2$. We assume that the turbulent wave field is statistically isotropic and we perform an average  over eight distinct directions, after rotation of the whole image so that $\mathbf k_2$ lies on the horizontal axis. It permits to compute $B_k(k_2,\mathbf k_3)$ with a better statistical convergence. Figure~\ref{fig10} displays $B_k(k_2,\mathbf k_3)$ for three $k_2$ and two forcing (weak on the left, strong on the right). The exacts solutions are plotted in black and the given $\mathbf k_2$ are represented by the black arrows. We observe a concentration of high correlation along the exact solutions presented before. As expected, due to the permutation equivalence, there is a vertical symmetry axis at $k_x/2\pi=0$ and $k_x/2\pi=k_2/2$. The magnitude of the correlation is about the same level of $10^{-1}$ observed for the frequency bicoherence $B_\omega(\omega_2,\omega_3)$. For the weak forcing, the repartition of the correlations shows a concentration of higher value along the 1D direction (horizontal axis with $k_y=0$), consistently with our conclusions coming from the analysis of $B_\omega(\omega_2,\omega_3)$. However, the observed resonances allow for a significant coupling of wavevectors with a quite large angle among each other notably at small $k_2$. The stronger forcing present the same general behavior with a much wider nonlinear spreading resulting, as expected, by the approximate resonances. The observed correlations also allow a much higher coupling between wave vector that are not collinear.

 \clearpage
 
\section{Discussion and concluding remarks} 

The investigations of 3-wave correlations confirm the existence of 3-wave non linear coupling in weak wave turbulence. In the weakest case, the observed correlations suggest that the dominant coupling occurs for wave triads that are close to collinear. If the dispersion relation takes the form of a pure power law $\omega\propto k^\alpha$ (with $\alpha\neq1$ for dispersive waves), no 3-wave resonance is possible for collinear wavevectors. In the case of gravity-capillary waves, the change of curvature allow for such resonances if waves have sufficiently separated frequencies. Our observations also underline the importance of approximate resonances permitted by nonlinear spectral widening. These approximate resonances permit notably 3-wave interactions among gravity waves that may induce a more efficient energy transfer among gravity waves than the 4-wave exact resonances. 

\begin{figure}[!htb]
\includegraphics[width=8cm]{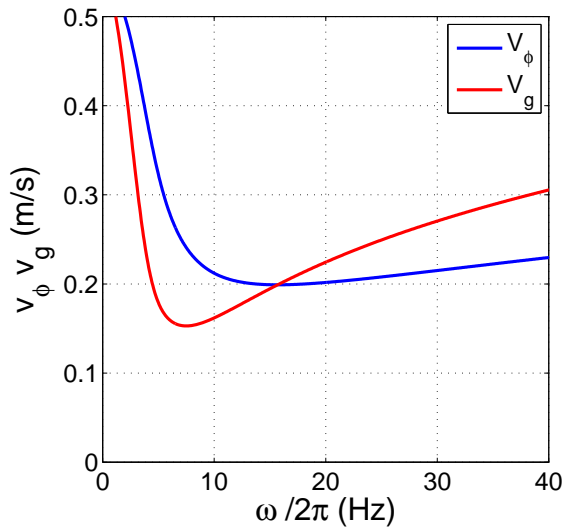}
\caption{Linear phase velocity $v_\phi=\frac{\omega}{k}$ and the linear group velocity $v_g=\frac{\partial \omega}{\partial k}$ in the conditions of our experiment. $v\phi$ present a minimum near $15$~Hz where the waves are non-dispersive. We note also that $v_\phi$ is quasi constant on the interval [10,30] Hz}
\label{fig11}
\end{figure}
The fact that the unidirectional resonances seem to dominate the correlations may be related to the fact that the phase and group velocities exhibit a minimum around the frequency of the gravity/capillary cross-over. The linear phase velocity $v_\phi=\frac{\omega}{k}$ and the linear group velocity $v_g=\frac{\partial \omega}{\partial k}$ are shown in fig.~\ref{fig11}. A direct effect of this double curvature is the presence of a local minimum on the phase velocity near $\omega /2\pi=15$~Hz and a minimum of the group velocity around 8~Hz.
In addition, we observe that $v_\phi$ remain roughly constant near this minimum in the interval $[10,30]$~Hz and the group velocity increases very slowly in the same range of frequencies: these waves are thus weakly dispersive. The combination of these two effects may increase significantly the efficiency of the 1D-interaction. Indeed, the waves packets moving with similar speeds in the same direction have a long time to interact. Then, for a weakly non-linear system, where the interaction time has to be long, 1D interaction are enhanced and finally, dominate. When the degree of non-linearity increases, the interaction time becomes shorter thus unidirectional interaction may not dominate as significantly. Thus, waves traveling with a slight angle may have time to interact and exchange similar amounts of energy. We also observed that the maximum level of correlation is constant for all our experiments (about $10^{-1}$). This constant value of the correlation suggests that the efficiency of the nonlinear interactions may be bounded. However, this possible saturation of the interactions remain to be confirmed with more strongly nonlinear experiments.    

The weak turbulence theory has been applied separately both to pure gravity surface waves and to capillary waves. Our observations show that at the crossover between both types of waves, the wave turbulence that is observed may be quite far from what is expected from the study of either regime separately. The striking example is the possibility to observe weak non linear interaction among collinear waves that may even be dominant due to the low dispersion of the waves in this range of frequencies. Another feature is the possibility to observe non local coupling between gravity waves and capillary waves as well as 3-wave coupling of gravity waves allowed by nonlinear spectral widening. Our observations illustrate the importance of approximate resonances permitted by non linearity that may change very significantly the dynamics of wave turbulence as compared to the case of exact resonances. If 3-wave coupling is permitted we should expect it to be much more efficient to transfer energy than the 4-wave coupling predicted in the case of exact resonances. It remains to investigate whether this 3-wave coupling is indeed observed in oceanic data. Another question is also the range of frequencies or wavelengths of gravity waves that may be impacted by this mechanism. Is it restricted to frequencies close to the gravity/capillary crossover or does it have an impact on much lower frequencies, for long waves ?

\begin{acknowledgements}
This project has received funding from the European Research Council (ERC) under the European Union's Horizon 2020 research and innovation programme (grant agreement No 647018-WATU). We thank Kronos Worldwide, Inc. for kindly providing us with the titanium oxide pigment Kronos 1001.
\end{acknowledgements}

\bibliography{biblio7}

\end{document}